\documentstyle[emulateapj,epsf]{article}

\newcommand {\simlt}{\lower.5ex\hbox{$\; \buildrel < \over \sim \;$}}
\newcommand {\simgt}{\lower.5ex\hbox{$\; \buildrel > \over \sim \;$}}
\newcommand {\boom}{{\sc Boomerang} }
\newcommand {\boomn}{{\sc Boomerang}}
\newcommand {\boomna}{{\sc Boomerang}/NA}
\slugcomment{}

\begin{document}

\title{Measurement of a Peak in the Cosmic Microwave Background\\
Power Spectrum from the North American test flight of \boomn}

\author{
P.D. Mauskopf\altaffilmark{1}, P.A.R. Ade\altaffilmark{2}, P. de Bernardis\altaffilmark{3}, J.J. Bock\altaffilmark{4,5}, 
J. Borrill\altaffilmark{6,7}, A. Boscaleri\altaffilmark{8}, B.P. Crill\altaffilmark{4},
G. DeGasperis\altaffilmark{9},
G. De Troia\altaffilmark{3}, P. Farese\altaffilmark{10}, P. G. Ferreira\altaffilmark{11,12},
K. Ganga\altaffilmark{13,14}, M. Giacometti\altaffilmark{3}, S. Hanany\altaffilmark{15}, 
V.V. Hristov\altaffilmark{4}, A. Iacoangeli\altaffilmark{3}, A. H. Jaffe\altaffilmark{6},
A.E. Lange\altaffilmark{4}, A. T. Lee\altaffilmark{16},
S. Masi\altaffilmark{3}, 
A. Melchiorri\altaffilmark{17}, F. Melchiorri\altaffilmark{3}, L. Miglio\altaffilmark{3},
T. Montroy\altaffilmark{10}, C.B. Netterfield\altaffilmark{18}, 
E. Pascale\altaffilmark{8}, F. Piacentini\altaffilmark{3}, P. L. Richards\altaffilmark{16},
G. Romeo\altaffilmark{19}, J.E. Ruhl\altaffilmark{10}, E. Scannapieco\altaffilmark{16},
F. Scaramuzzi\altaffilmark{20}, R. Stompor\altaffilmark{16} and N. Vittorio\altaffilmark{9}
}


\affil{
 $^1$ Dept. of Physics and Astronomy, University of Massachusets, Amherst, MA, USA\\
 $^2$ Queen Mary and Westfield College, London, UK \\
 $^3$ Dipartimento di Fisica, Universita' La Sapienza, Roma, Italy \\
 $^4$ California Institute of Technology, Pasadena, CA, USA \\
 $^5$ Jet Propulsion Laboratory, Pasadena, CA, USA \\
 $^6$ Center for Particle Astrophysics, University of California, Berkeley, CA, USA \\
 $^7$ National Energy Research Scientific Computing Center, LBNL, Berkeley, CA, USA \\
 $^8$ IROE-CNR, Firenze, Italy \\
 $^9$ Dipartimento di Fisica, Universita' di Roma Tor Vergata, Roma, Italy \\
 $^{10}$ Dept. of Physics, Univ. of California, Santa Barbara, CA, USA \\
 $^{11}$ CENTRA, IST, Lisbon, Portugal \\
 $^{12}$ Theory Division, CERN, Geneva, Switzerland \\
 $^{13}$ Physique Corpusculaire et Cosmologie, College de France, 11 Place
Marcelin Berthelot, 75231 Paris Cedex 05, France\\
 $^{14}$ IPAC, Pasadena, CA, USA\\
 $^{15}$ Dept. of Physics, University of Minnesota, Minnealpolis, MN, USA \\
 $^{16}$ Dept. of Physics, University of California, Berkeley, CA, USA \\
 $^{17}$ Dept. de Physique Theorique, Universite de Geneve, Switzerland \\
 $^{18}$ Depts. of Physics and Astronomy, University of Toronto, Canada \\
 $^{19}$ Istituto Nazionale di Geofisica, Roma, Italy \\
 $^{20}$ ENEA, Frascati, Italy\\
}

\begin{abstract}

We describe a measurement of the angular power spectrum of anisotropies in the Cosmic
Microwave Background (CMB) from $0.3^{\circ}$ to $\sim 10^{\circ}$
from the North American test flight of the \boom
experiment. \boom is a balloon-borne telescope with a bolometric
receiver designed to map CMB anisotropies on a Long Duration Balloon
flight.  During a 6-hour test flight of a prototype system
in 1997, we mapped $> 200$ square 
degrees at high galactic latitudes in two bands centered at 90 and 150~GHz
with a resolution of 26 and 16.6 arcmin FWHM respectively. Analysis of the
maps gives a power spectrum with a peak at angular scales of $\sim 1$ degree
with an amplitude $\sim 70 \mu$K$_{CMB}$.
\end{abstract}

\keywords{cosmology: Cosmic Microwave Background, anisotropy, measurements,
power spectrum}

\section{Introduction}

Measurements of Cosmic Microwave Background (CMB) anisotropies
have the potential to
reveal many of the fundamental properties of the universe
(e.g. see \cite{review} and references therein). 
Since the measurement of the large scale
anisotropies by COBE-DMR (\cite{cobe2}, \cite{bennett}, etc.), many 
ground-based and balloon-borne experiments have
continued to develop the technology necessary to produce
accurate measurements of CMB structure on smaller angular
scales.  Recent results from these experiments have shown
evidence for the existence of a peak in the power spectrum
of CMB fluctuations at a multipole, $\ell \sim 200$
(\cite{cbn}, \cite{QMAP}, \cite{MSAMI}, \cite{TOCO97}, 
\cite{pythonv}, \cite{MAT}).  In this paper, we present results from the 
test flight of \boom that constrain the position and amplitude of this peak.
In fact, the results from this data set alone can be used to
constrain cosmological models and provide evidence for a flat universe,
$\Omega = \Omega_M + \Omega_{\Lambda} \simeq 1$ (see the companion
paper, Melchiorri, et al. 1999).

\boom is a millimeter-wave telescope and receiver system designed for
a Long Duration Balloon (LDB) flight
from Antarctica of up to two weeks duration.  The data
presented here are from a prerequisite test flight of the payload
in August, 1997, which lasted 6 hours at float altitude.
A prototype of the LDB focal plane was used to test
the experimental strategy of making a map and measuring the angular
power spectrum of CMB anisotropy with a slowly-scanned telescope and
total-power bolometric receiver.

\section{Instrumentation}

\boom employs
bolometric detectors cooled to 0.3~K, coupled to an off-axis
telescope and operated as total power radiometers.
Electronic modulation and synchronous demodulation of the bolometer
signals provide low frequency stability of the detector system.
The optical signal is modulated by slowly scanning the telescope
($\sim 1$ deg/s) in azimuth.
A description of the instrument can be found in \cite{b99tech}.

The \boom telescope consists of three off-axis, aluminum mirrors:
an ambient temperature 1.3~m primary mirror and two smaller ($15$~cm)
2~K reimaging mirrors.
The telescope is shielded from ground radiation by extensive
low emissivity baffles.
Incoming radiation is reflected by the primary mirror 
into the entrance window of the cryostat and reimaged
by the 2~K optics onto the 0.3~K focal plane.  The tertiary mirror
is positioned at an image of the primary mirror and forms a throughput-limiting
Lyot stop illuminating the central 85~cm of the primary mirror.  The inside of
the 2~K optics box has a thick coating of absorbing material (\cite{bockblack})
to intercept stray light.  A calibration lamp mounted in the Lyot stop of the
optical system is pulsed for $\sim$ 0.5 s every 15 minutes to produce a stable,
high signal-to-noise ratio ($\simgt 1000$) calibration transfer.

For the test flight,
the focal plane contained silicon nitride micromesh bolometers
(\cite{bolo1}, \cite{bolo2}) 
fed by multi-mode feed horns with channels at 90 and 150~GHz.
Fourier transform spectroscopy of each channel determined the
average band centers and effective band widths
for sources with a spectrum of CMB anisotropy (Table 1).
A long duration $^3He$ cryostat (\cite{cryo1}, \cite{cryo2}) 
maintained the focal plane at $0.285 \pm 0.005$~K during the flight.
The bolometers were biased with an AC current and
the signals were demodulated, amplified and passed through a 4-pole
Butterworth low-pass filter
with a cut-off frequency of $20$~Hz for the 150~GHz channels and $10$~Hz 
for the 90~GHz channels.  The bolometer DC levels were removed with a
single-pole, high-pass filter with a cut-on frequency of $16$~mHz and
re-amplified before digital sampling at 62.5~Hz with 16 bit resolution.

The attitude control system
included a low-noise two-axis flux-gate magnetometer, 
a CCD star camera and three orthogonal (azimuth, pitch, roll) 
rate gyroscopes.
The azimuth gyroscope drove the azimuth feedback loop, which 
actuated two torque motors, one of which drove a flywheel while the other
transferred angular momentum to the balloon flight line.
The gyroscope was referenced every
scan to the magnetometer azimuth to remove long time scale drifts.
Pendulations of the payload were reduced by a oil-filled damper mounted
near the main bearing and
sensed by the pitch and roll gyroscopes outside of the control loop.
The CCD frame had $480 \times 512$ pixels, with a resolution of  
0.81 arcmin/pixel (azimuth) and 0.65 arcmin/pixel (elevation). 
The limiting magnitude of stars used for pointing  
reconstruction was  $m_V \sim 5$. 
The coordinates of the two brightest stars in each CCD frame were identified by 
on-board software and recorded at 5~Hz,
for post-flight attitude reconstruction.

The primary mirror and the receiver were mounted on an inner frame 
which could be rotated to position the elevation boresight of the telescope between
$35^{\circ} - 55^{\circ}$.  During observations, the inner
frame was held fixed while the entire gondola slowly scanned in
azimuth.  The scan speed was set to the maximum allowed by the beam size and
bolometer time constants.
The window functions for the 90 and 150~GHz channels in this flight
were rolled off at $\ell \simgt 300$ and $500$, respectively, by the 26
and 16.5 arcmin beams and were limited to $\ell \simgt 20$ by the system 1/f noise.

\section{Observations} 
 
The instrument was launched at 00:25 GMT on August 30, 1997 from the  
National Scientific Balloon Facility in Palestine, Texas. 
Observations started at 3:50 GMT and continued until the flight was
terminated at 9:50 GMT. Sunset was at 01:00 GMT and the moon rose at
10:42 GMT.  The average altitude at float 
was 38.5 km.  We observed in three modes:
$\sim 4.5$ hours of CMB scans, 40 minutes of calibration scans on Jupiter,
and four ten-minute periods of full-sky rotations of the gondola.
The CMB observations consisted of smoothed triangle wave scans in azimuth
with a peak-to-peak amplitude of 40 degrees centered on due South at
an elevation angle of $45$~degrees.
The scan rate was $\sim 2.1$ deg/s in azimuth ($1.4$~deg/s on the sky)
in the linear portion of the scan (63$\%$ of the period).
These scans, combined with the earth's rotation,
covered a wide strip of sky, from $-73^{\circ}< {\rm RA} < 23^{\circ}$
and $-20^{\circ} < {\rm DEC} < -16^{\circ}$.

\section{Data reduction}

We obtain a first order reconstruction of the attitude from the  
elevation encoder and magnetometer data.  Using this, the stars recorded
by the CCD camera are identified. The data from the gyroscopes are
used to interpolate between star identifications.
The offset between the microwave telescope and the 
star camera is measured from the observation of 
Jupiter. The attitude solution obtained in 
this way is accurate to $\sim$ 1 arcmin rms. 

The results presented in this paper are primarily from the highest
sensitivity 150~GHz channel.  Data from a 90~GHz channel are presented
as a systematic check.
The data at float consist of $\sim$ 1.4 million samples for each bolometer.
These data are searched for large ($>5 \sigma$)
deviations such as cosmic ray events and radio frequency interference
and flagged accordingly.
Flags are also set for events in the auxiliary data such as calibration
lamp signals and elevation changes, as well as for the different
scan modes.  Unflagged data, used for the CMB analysis,  
are 54$\%$ of the total at 150~GHz and 60$\%$ of the total at
90~GHz.  We deconvolve the transfer functions of the readout electronics and
the bolometer thermal response from the data, and apply a flat phase numeric  
filter to reduce high frequency noise and slow drifts.
The time constants of the bolometers given in Table 1 are measured in
flight from the response to cosmic rays and to fast (18 deg/s) scans over 
Jupiter.
 
\smallskip
\begin{center}
{\footnotesize
\vbox{Table 1. \boom 1997 instrument parameters}
\smallskip
\begin{tabular}{cccccc}
\tableline\tableline
\noalign{\vskip .4em}
$\nu_0$ & $\Delta \nu$ & FWHM & $\Omega$ & $\tau_b$ & NET$_{\rm CMB}$\\
(GHz) & (GHz) & ($'$) & $10^{-5}$ sr & (ms) & ($\mu$K$/\sqrt{\rm Hz}$) \\\hline
\noalign{\vskip .4em}

96 & 33 & 26 & $6.47 \pm 0.27$ & $71 \pm 8$ & $400$\\

153 & 42 & 16.5 & $2.63 \pm 0.10$ & $83 \pm 12$ & $250$ \\\hline

\noalign{\vskip .4em}

\end{tabular}}
\end{center}

The primary calibration is obtained from 
scans of the telescope over Jupiter, which were performed 
from 3:32 U.T. to 3:53 U.T. (1.2 deg/s azimuth scans). 
Jupiter was also re-observed later during regular CMB scans 
(from 5:58 to 6:18 U.T.) and during full sky rotations.
Jupiter was rising during the first 
observation (from an elevation of 36.9$^{\circ}$ to 38.2$^{\circ}$) 
and setting during the second one (from an elevation 
of 40.1$^{\circ}$ to 39.0$^{\circ}$).  The deconvolved and filtered data
from these scans are triangle-interpolated on a regular grid centered on the
optical position
of Jupiter to make a beam map. The beams are symmetric with minor and
major axes equivalent within $5$\%.  Solid angles given in Table 1 are 
computed by integration of the interpolated data.  The beam map is used to 
derive the window function, $W_{\ell}$, and an overall normalization.

We use a brightness 
temperature of $T_{\rm eff} = 173$~K for both the 90 and 150~GHz bands 
(\cite{Ulich}, \cite{Griffin}, \cite{Cole}) and assume an uncertainty of
$5\%$ (\cite{MSAM}).
We also assign a $5\%$ error in the conversion from brightness temperature
to CMB temperature due to statistical noise in the measurements of the 
band pass.
Errors in the determination of the receiver transfer function are largest at
the highest temporal frequencies where signals are attenuated by the bolometer
time constants.  These errors affect the calibration
of the window function and the CMB power spectrum measurements at the largest
multipoles.
These should cancel because the beam maps of Jupiter are made from
scans at the same scan speed as the CMB scans.  In simulations, a 15$\%$
error in the determination of the bolometer time constant produces a
maximum error 
$<1.5\%$ in the normalization and $\simlt 6 \%$ in the value of $W_{\ell}$ at
$\ell < 300$ near scan turnarounds where the scan speed was only 1 deg/s.
The final precision of the calibration including errors
in the temperature of Jupiter, $T_{\rm eff}$, the transfer function used for
deconvolution, the measurement of the filter pass bands
and the beam solid angle is 8.1$\%$ and 8.5$\%$ at 150~GHz and 90~GHz, 
respectively.
 
Signals from the internal calibration lamp are used to correct for a slow linear
drift in detector responsivity during the flight, mainly due to variation (0.29-0.28~K)
in the base temperature of the fridge.  The change in response to the 
calibration lamp from the  beginning to the end of the flight is
$+9\%, -1\%$ at 90~GHz, and $+8\%, -1.5\%$ at 150~GHz. 

\section{Map Making and Power Spectrum Estimates} 

Current and future CMB missions require new methods of analysis
able to incorporate the effects of correlated instrument
noise and new implementations capable of processing large data
sets (Bond et al. 1999).
The analysis of the \boom test flight data provides a test of these
methods on a moderate size ($\sim 25,000$ pixels) data set.
The calibrated time
stream data are processed to produce a pixelized map, and from this a
measurement of the angular power spectrum, using
the MADCAP software package of Borrill (1999a, 1999b) (see
http://cfpa.berkeley.edu/$\sim$borrill/cmb/madcap.html)
on the Cray T3E-900 at NERSC and the Cray T3E-1200 at CINECA.

Noise correlation functions are estimated from the time stream under
the assumption that in this domain the signal is small compared to the
noise. For the 150~GHz channel, the maximum likelihood map with 23,561,
1/3-beam sized ($6.9'$) pixels is calculated from
the noise correlation function, the bolometer signal (excluding
flagged data and all data within 2 degrees of Jupiter) and the
pointing information using the procedures described in Wright (1996),
Tegmark (1997) and Ferreira \& Jaffe (1999).
The maximum likelihood power spectrum is estimated from
the map using a Newton-Raphson iterative maximization of the $C_\ell$
likelihood function following Bond, Jaffe \& Knox (1998) (BJK98) and
Tegmark (1998).
The power spectrum analysis requires approximately 12 hours on sixty-four
T3E-900 processors.

\smallskip
{\footnotesize
\vbox{\sc Table 2: Orthogonalized power spectrum measurement
from \boomna 150~GHz channel map with with 6 arcminute pixels.}}
\begin{center}
\smallskip
{\footnotesize
\begin{tabular}{llcc}
\tableline\tableline
\noalign{\vskip .4em}
$\ell_{\rm eff}$ & $[\ell_{\rm min}, \ell_{\rm max}]$ & $\Delta T$ & 
$\ell(\ell+1)C_{\ell}/2\pi$\\
 &  & ($\mu$K)$^a$ & ($\mu$K$^2$)$^b$\\\hline
\noalign{\vskip .4em}

58 & $\left[25,75\right]$ &  $29^{+13}_{-11}$  & $850^{+900}_{-540}(\pm660)$ \\
102 & $\left[76,125\right]$ &  $49^{+9}_{-9}$  &  $2380^{+990}_{-780}(\pm860)$ \\
153 & $\left[126,175\right]$ &  $67^{+10}_{-9}$  &  $4510^{+1380}_{-1140}(\pm1250)$ \\
204 & $\left[176,225\right]$ &  $72^{+10}_{-10}$  &  $5170^{+1500}_{-1320}(\pm1410)$ \\
255 & $\left[226,275\right]$ &  $61^{+11}_{-12}$  &  $3700^{+1500}_{-1300}(\pm1390)$ \\
305 & $\left[276,325\right]$ &  $55^{+14}_{-15}$  &  $3070^{+1680}_{-1530}(\pm1570)$ \\
403 & $\left[326,475\right]$ &  $32^{+13}_{-22}$  &  $1030^{+1020}_{-900}(\pm1180)$ \\
729 & $\left[476,1125\right]$ &  $<130$  &  $250^{+16500}_{-250}(\pm2710)$ \\
\hline

\noalign{\vskip .4em}
\end{tabular}}
\end{center}
{\footnotesize
\vbox{$^a$ Errors from 68\% confidence intervals}
\vbox{$^b$ Errors from 68\% confidence intervals with $\pm 1 \sigma$ assuming
Gaussian likelihood function in parentheses}}

\medskip 
\centerline{\vbox{\epsfxsize=7.5cm\epsfbox{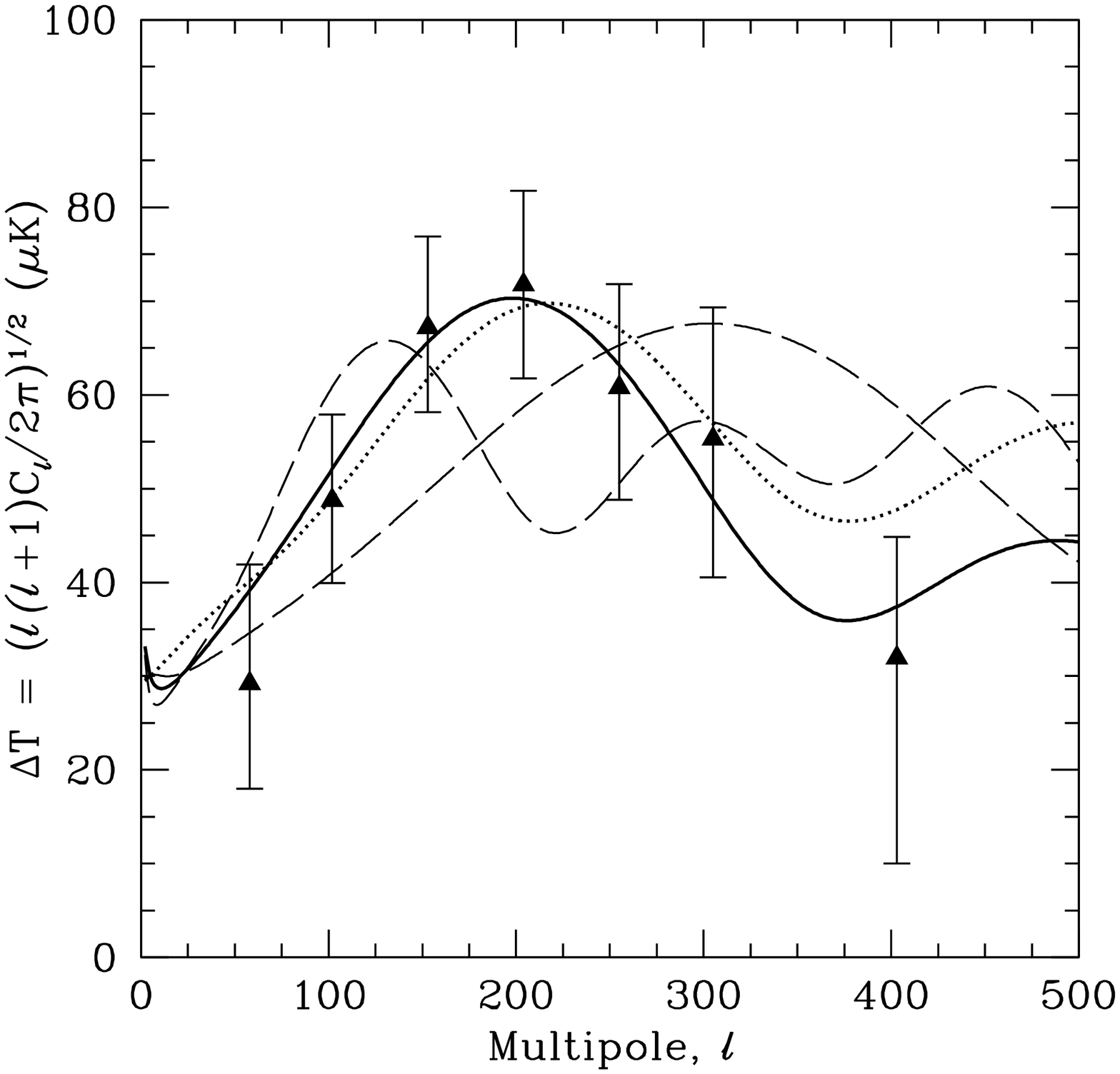}}}
{\small F{\scriptsize IG}.~1.--- 
Power spectrum of the \boom 150~GHz map with 6
arcminute pixelization.  The solid curve is a marginally closed
model with ($\Omega_b,\Omega_M,\Omega_{\Lambda},n_S,h) =
(0.05, 0.26, 0.75, 0.95, 0.7)$. The dotted curve
is standard CDM with $(0.05, 0.95, 0.0, 1.0, 0.65)$.
Dashed curves are open and closed models with fixed
$\Omega_{\rm total} = 0.66$ and $1.55$, respectively (see the companion
paper, \cite{parameters}).
\label{fig:powspec}} \medskip

Due to finite sky coverage, individual $C_\ell$
values are not independent, and we estimate the power in eight bins over
$25<\ell<1125$, binning more finely at low $\ell$ than high.
In each bin we calculate the maximum likelihood amplitude of a flat
power spectrum in that bin, i.e. $\ell(\ell+1)C_{\ell} = $constant (Table 2).
Given our insensitivity to low $\ell$ signal, and in order to
ameliorate the effects of low-frequency noise (which
translates into low-$\ell$ structure for this scanning strategy), we
marginalize over power at $\ell<25$ and diagonalize the bin
correlation matrix using a variant of techniques discussed in BJK98. 
Bin to bin signal correlations are small but non-negligible, with each
top-hat bin anticorrelated with its nearest neighbors at approximately
the 10\% level.

We calculate likelihood
functions for each bin using the offset log-normal distribution
model of BJK98.  We quote errors for the
power spectrum from $\pm 1 \sigma$ errors assuming
simple Gaussian likelihood functions and from 68\% confidence intervals
for the full likelihood functions.  Plots of the likelihood functions
are presented in the companion paper (Melchiorri, et al. 1999) and
full spectral data including information on the shape of the likelihood
function (Bond, Jaffe \& Knox 1998b) and appropriate window and filter
functions (Knox 1999) will be available at the \boom web site
(http://boom.physics.ucsb.edu or http://oberon.roma1.infn.it/boom).
In Figure 1, we show the calculated power spectrum for the 150 GHz
channel 6 arcminute pixel map.

\section{Systematics checks and Foregrounds}

The scan strategy and analysis allow for a variety of checks for
systematic effects in the data. For all systematic tests we produce small
($\sim 4000$ pixel) maps with $16^{\prime}$ pixels which can be generated and
analyzed quickly ($\sim$ 30 minutes on four T3E-900 processors).
Monte-Carlo simulations show the effects of the 16$'$ pixelization are
negligible for values of $\ell < 400$.  Compared to the level of statistical
noise, we find that the derived power spectrum from the 150~GHz
channel is insensitive to (i) the size of the excluded region around Jupiter,
(ii) the choice of spectral shape within the multipole bins,
and (iii) the method of marginalizing over low $\ell$ signals.

We also analyze maps made from combinations of data that we expect to
produce null power spectra in the absence of spurious sources of
noise power: (i) data from a dark detector and (ii) the difference between
left and right-going scans at 150 GHz.  In each case (Table 3),
we find a power spectrum consistent with zero.

\smallskip
{\footnotesize
\vbox{Table 3. Power Spectra, $\ell(\ell+1)C_{\ell}/2\pi$, from \boom
systematic tests.
Error bars are $\pm 1 \sigma$, assuming Gaussian likelihood functions.
Unlike the data in Table 2 and Figure 1, these spectra are calculated with 16
arcminute pixels and have not been orthogonalized.  The (L-R) difference
analysis in column 5 has not been corrected for the effects of correlated
noise, which leads to negative power estimates in the high $\ell$ bins.}}
\begin{center}
{\footnotesize
\begin{tabular}{lcccc}
\tableline\tableline
\noalign{\vskip .4em}
$[\ell_{\rm min}, \ell_{\rm max}]$ & 150 GHz & 90 GHz & Dark &
150 (L-R)/2\\\hline
 & ($\mu$K$^2/100$) & ($\mu$K$^2/100$) & ($\mu$K$^2/100$) & 
($\mu$K$^2/100$)\\\hline
\noalign{\vskip .4em}

$\left[25,75\right]$ & $10\pm7$ &
$16\pm12$ & $-1\pm1$  &  $4\pm7$ \\

$\left[76,125\right]$ & $23\pm9$ &
$25\pm13$  &  $1\pm1$ & $-9\pm4$ \\

$\left[126,175\right]$ & $46\pm13$ &
$50\pm20$  &  $1\pm2$ & $-7\pm9$ \\

$\left[176,225\right]$ & $50\pm14$ &
$53\pm25$  &  $6\pm4$ &  $12\pm17$ \\

$\left[226,275\right]$ & $29\pm13$ &
$21\pm29$  &  $-1\pm5$ &  $-22\pm18$ \\

$\left[276,325\right]$ & $23\pm15$ &
$5\pm40$  &  $2\pm7$ &  $-30\pm26$ \\

$\left[326,475\right]$ & $8\pm12$ &
$3\pm48$  &  $7\pm5$ &  $-78\pm21$ \\

$\left[476,1125\right]$ & $2\pm27$ & 
$2\pm253$  &  $9\pm12$ & $-169\pm52$ \\\hline

$\left[25,475\right]$ & $31\pm5$ &
$25\pm7$ & $-1\pm1$  &  $-4\pm2$ \\

\noalign{\vskip .4em}

\end{tabular}}
\end{center}

The sky strip observed is extended in Galactic latitude from $b \sim
15^\circ $ to $b \sim 80^\circ $. From IRAS/DIRBE map extrapolation
(Schlegel et al. 1998) dust emission is not expected to produce
significant contamination in this region of the sky. We find an
amplitude for a flat power spectrum from $25<\ell<400$ of $\ell
(\ell+1)C_{\ell}/2\pi = (3580\pm706) \mu$K$^2$ and $(2982\pm702) \mu$K$^2$
from the half of the map near the galactic
plane ($15^{\circ}<b<45^{\circ}$) and the  
higher-latitude half ($45^{\circ}<b<80^{\circ}$), respectively.  Because
of sky rotation, this also corresponds to separately analyzing the
data from the first half and second half of the flight.

Jupiter is more than 1000 times brighter than degree-scale anisotropy in
the CMB and is in the middle of the lowest latitude half of the map.
We test for sidelobe contamination by
varying the size of the circular avoidance zone around Jupiter from $\phi = 0.5-3$
degrees for the CMB maps,
and find no significant change in the power spectrum as long as we
remove data within $\sim 1$ degree.

\section{Discussion and Conclusions} 
 
Data obtained in $\sim 4.5$ hours of CMB scans during the \boom test
flight have been analyzed to produce maps of the millimeter-wave sky
at 90 and 150 GHz,   with 26 and 16 arcmin FWHM resolution, respectively.  
The instrument was calibrated using observations 
of Jupiter.  The experiment employed a new
scan strategy and detector technology designed to give maximum
coverage of angular scales.  The data analysis 
implements new techniques for making maximum likelihood maps from
low signal-to-noise time stream data over large numbers of pixels.
The power spectrum $C_\ell$ obtained 
from the maps extends from $\ell \sim 50$ to $\ell \sim 800$ and 
shows a peak at $\ell \sim 200$.
  
Since the \boomna
test flight in August, 1997, we have obtained
$> 200$ hours of data from the LDB flight of \boom (\boom/LDB), carried 
out in Antarctica at the end of 1998.  The focal plane for the LDB flight
contained 2, 6, and 3 detectors at 90, 150, and 240~GHz,
each with better sensitivity to CMB, faster time constants, and lower 1/f noise
than the best channel in the \boomna flight, as well as 3 detectors at 400~GHz
to provide a monitor of interstellar dust and atmospheric emission.
The results from this flight will be reported elsewhere. 
 
\acknowledgments 
 
We thank the staff of the NASA NSBF for their excellent support 
of the test flight.  We are also grateful for the help and support
of a large group of students and scientists (a list of contributors
to the \boom experiment can be found at 
http://astro.caltech.edu/$\sim$bpc/boom98.html).
The \boom program has been supported by Programma Nazionale Ricerche
in Antartide, Agenzia Spaziale Italiana and University of Rome La Sapienza
in Italy; by NASA grant numbers NAG5-4081, NAG5-4455, NAG5-6552, NAG53941,
by the NSF Science \& Technology
Center for Particle Astrophysics grant number SA1477-22311NM under
AST-9120005, by NSF grant 9872979, and by the NSF Office of Polar Programs 
grant number OPP-9729121 in the USA; and by PPARC in UK.
This research also used resources of the National Energy Research
Scientific Computing Center, which is supported by the Office of
Science of the U.S. Department of Energy under Contract
No. DE-AC03-76SF00098.  Additional computational support for the
data analysis has been provided by CINECA/Bologna.

\end{document}